\documentclass[twocolumn]{aastex63}

\usepackage{times,color,natbib,url,amsmath}
\usepackage{graphicx,float,psfrag}
\usepackage{grffile}
\usepackage{tabularx}
\usepackage{latexsym,amsmath,amssymb}
\usepackage{natbib}
\usepackage{verbatim}
\usepackage{multirow}
\usepackage{color}
\usepackage[hang]{footmisc}
\usepackage{soul,xcolor}

\bibpunct{(}{)}{;}{a}{}{,}

\newcommand {\einstein} {{\it Einstein}}

\newcommand {\xmm} {\textsl{XMM-Newton}}

\newcommand {\swift} {\textsl{Swift}}
\newcommand {\rxte} {\textsl{RXTE}}
\newcommand {\fermi} {\textsl{Fermi}}
\newcommand {\nustar} {\textsl{NuSTAR}}

\newcommand {\nicer} {\textsl{NICER}}

\def\deg{$^\circ$}
\def \rsun {\ifmmode$R$_{\odot}\else R$_{\odot}$}

\def \hcm {\hbox {\ifmmode $ atoms cm$^{-2}\else atoms cm$^{-2}$\fi}}

\def\approxgt{\mathrel{\hbox{\rlap{\lower.55ex \hbox {$\sim$}}
        \kern-.3em \raise.4ex \hbox{$>$}}}}
\def\approxlt{\mathrel{\hbox{\rlap{\lower.55ex \hbox {$\sim$}}
        \kern-.3em \raise.4ex \hbox{$<$}}}}

\def \arcmin {\hbox{$^\prime$}}

\newcommand\T{\rule{0pt}{2.6ex}}
\newcommand\B{\rule[-1.2ex]{0pt}{0pt}}

\def \src {1E~2259+586}

\begin{document}
\setstcolor{red}

\title{\rm \uppercase{A radiatively-quiet glitch and anti-glitch in
    the magnetar \src}}

%
% Authors
%% Authors with the same affiliation can be grouped in a single
%% \author and \affil call.
\author{George~Younes}
\affiliation{Department of Physics, The George Washington University, Washington, DC 20052, USA, gyounes@gwu.edu}
\affiliation{Astronomy, Physics and Statistics Institute of Sciences (APSIS), The George Washington University, Washington, DC 20052, USA}
\author{Paul~S.~Ray}
\affiliation{Space Science Division, U.S. Naval Research Laboratory, Washington, DC 20375, USA}
\author{Matthew~G.~Baring}
\affiliation{Department of Physics and Astronomy, Rice University, MS-108, P.O. Box 1892, Houston, TX 77251, USA}
\author{Chryssa~Kouveliotou}
\affiliation{Department of Physics, The George Washington University, Washington, DC 20052, USA, gyounes@gwu.edu}
\affiliation{Astronomy, Physics and Statistics Institute of Sciences (APSIS), The George Washington University, Washington, DC 20052, USA}
\author{Corinne~Fletcher}
\affiliation{Science and Technology Institute, Universities Space Research Association, Huntsville, AL 35805, USA}
\author{Zorawar~Wadiasingh}
\affiliation{Astrophysics Science Division, NASA Goddard Space Flight Center, Greenbelt, MD 20771}
\author{Alice~K.~Harding}
\affiliation{Astrophysics Science Division, NASA Goddard Space Flight Center, Greenbelt, MD 20771}
\author{Adam~Goldstein}
\affiliation{Science and Technology Institute, Universities Space Research Association, Huntsville, AL 35805, USA}

\begin{abstract}

We report on the timing and spectral properties of the soft X-ray
emission from the magnetar \src\ from January 2013, $\sim 8$ months
after the detection of an anti-glitch, until September 2019, using the
Neil Gehrels \swift\ and \nicer\ observatories. During this time span, we detect two timing discontinuities. The first, occurring around 5 years after the April 2012
anti-glitch, is a relatively large spin-up glitch with a fractional
amplitude $\Delta\nu/\nu=1.24(2)\times10^{-6}$. We find no evidence
for flux enhancement or change in the spectral or pulse profile
shape around the time of this glitch. This is consistent with the
picture that a significant number of magnetar spin-up glitches are
radiatively-quiet. Approximately 1.5~years later in April 2019, \src\
exhibited an anti-glitch with spin-down of a fractional amplitude
$\Delta\nu/\nu=-5.8(1)\times10^{-7}$; similar to the fractional change
detected in 2012. We do not, however, detect any change to the pulse-profile shape or increase in the rms
pulsed flux of the source, nor do we see any possible bursts from its
direction around the time of the anti-glitch; all of which occurred during the 2012 event. Hence,
similar to spin-up glitches, anti-glitches can occur silently. This may suggest that these phenomena originate in the neutron star
interior, and that their locale and
triggering mechanism do not necessarily have to be connected to the magnetosphere. Lastly, our observations suggest
that the occurrence rate of spin-up and spin-down glitches is about the same in
\src, with the former having a larger net fractional change.

\end{abstract}

\section{Introduction}
\label{Intro}

Magnetars represent a subset of the isolated neutron star (ISN)
family with a unique set of observational properties. Most show long
spin periods ($P\sim2$--$12$~s) and large spin-down rates ($\dot{P}
\sim10^{-13}$--$10^{-10}$~s~s$^{-1}$), implying large surface dipole
magnetic field strengths of the order of $\sim10^{14}$~G, and young
spin-down ages with an average of a few thousand years. Magnetars are
usually observed as hot thermal X-ray emitters with surface blackbody
temperatures of $kT\sim0.5$~keV, and as bright persistent X-ray sources
with $L_{\rm X}\sim10^{33}$--$10^{36}$~erg~s$^{-1}$, exceeding their
corresponding rotational energy losses ($|\dot{E}|\propto\dot{P}/P^3$).
Hence, unlike their less-magnetic cousins, rotation-powered
pulsars (RPPs), magnetars are believed to be powered through the decay
of their large inferred surface and internal magnetic fields \citep[see, e.g.,][for reviews]{mereghetti15:mag,
  turolla15:mag,kaspi17:magnetars}.

A defining trait of the magnetar class is their recurring variability
observed on broad time-scales. They randomly enter burst-active
episodes where they emit tens to hundreds of short ($\sim0.2$~s),
bright ($L_{\rm peak}\sim10^{40}$~erg~s$^{-1}$), hard X-ray bursts
over the course of days to months. Coincident with these bursting
episodes, an increase in their persistent X-ray emission by factors of
few to a thousand is most often observed. At the same time, the
persistent emission of magnetars undergoes changes to its spectral and
temporal properties, which often recover exponentially back
to quiescence over weeks to months time-scales \citep[e.g., ][]{camero14MNRAS,
  scholz14ApJ:1822,younes17:1935,cotizelati18MNRAS}. We note that the
above canonical characteristics are no longer restricted to typical (high dipolar $B$)
magnetars and have been recently observed from low$-B$ magnetars
\citep{rea10Sci:0418}, central compact objects \citep{borghese18MNRAS:rcw103},
and high$-B$ RPPs \citep{archibald16:j1119,gogus16:j1119}.

\src\ was discovered with the \einstein\ telescope in
the supernova remnant (SNR) G109.1$-$1.0 \citep{fahlman81Natur}. It has a
spin period of $P\approx7$~s and a spin-down rate of
$\dot{P} = 4.8\times10^{-13}$~s~s$^{-1}$, implying a
  surface polar field of $B\sim 1.2\times10^{14}$~G and a spin
down age of $P/(2{\dot P})\sim230$~kyr. In 2002, \src\ entered a
burst-active episode during which \rxte\ detected $\sim80$
magnetar-like bursts \citep{kaspi03ApJ:2259}. This discovery sealed
the earlier results by \citet{gavriil02Natur:AXPs} on the unification
of two classes of isolated neutron stars (INS), the Soft Gamma
Repeaters (SGRs) and the Anomalous X-ray Pulsars (AXPs), under the
magnetar umbrella. The source has been regularly monitored in the soft
X-ray band, first with \rxte, followed with \swift. Apart from the
outburst, \src\ has shown a relatively high level of
spectral and timing stability since its monitoring started in 1996,
the latter interrupted by discontinuities at a rate of about 1 every 6
years \citep{dib14ApJ}.

In 2012, \src\ entered an active episode where it showed both bursting
activity as well as an increase in its X-ray flux accompanied by
hardening of the spectrum \citep{archibald13Natur}. During this
episode, the source exhibited two discontinuities in its timing
behavior. The first, occurring at outburst onset, can only be
interpreted as an abrupt spin-down or anti-glitch event: a sudden
decrease in spin frequency with a fractional change of
$\Delta\nu/\nu\sim-3\times10^{-7}$. The second, could either be due to a regular spin-up glitch or 
another spin-down event, depending on the timing model \citep[see also][]{hu14ApJ:2259}. Spin-down 
glitches are exceptionally rare and have so far never been
reported from any RPP, which have collectively shown hundreds of
spin-up glitches
\citep[e.g.,][]{espinoza11MNRAS:glitches}\footnote{http://www.jb.man.ac.uk/pulsar/glitches/gTable.html\\https://www.atnf.csiro.au/research/pulsar/psrcat/glitchTbl.html}. Apart
from the 2012 event, another candidate spin-down glitch from \src\
occurred in 2009 \citep{icdem11MNRAS:1626,dib14ApJ}, which was also
accompanied by an elevated flux level from the source.  Note also that an
anti-glitch was reported from the magnetar 1E~1841$-$045 in archival
\rxte\ data \citep{sasmaz14MNRAS:1841}, however, analysis by a
different team of the same dataset returned a null result
\citep{dib14ApJ}.

In this paper, we report on our timing and spectral analyses of over
6.5~years of \swift\ and 9 months of \nicer\ data of the magnetar \src. During this
span, the source has shown a relatively large spin-up glitch and an
anti-glitch with a similar fractional change to the one detected in
2012. Both events are, however, radiatively quiet, contrasting the
2012 anti-glitch.  The observations
and data reduction are presented in Section~\ref{obs}. We summarize
our results in Section~\ref{res} and in Section~\ref{discuss}, we discuss
the implications of our discovery, focusing on the anti-glitch triggering locale, i.e., internal vs external to the neutron star.

\section{Observations and data reduction}
\label{obs}

\src\ was observed with \nicer\ on a bi-weekly basis starting on 2019
March 17, as part of our magnetar monitoring program. \nicer\ is a
non-imaging X-ray timing instrument, sensitive to photons in the
energy range 0.2--12~keV \citep{gendreau16SPIE}. It consists of 56
coaligned X-ray concentrating optics, covering a 30~arcmin$^2$ field
of view, providing a collecting area of 1900~cm$^2$ at 1.5 keV
\citep{lamarr16SPIE}. We processed \nicer\ data using NICERDAS version
6, as part of HEASOFT version 6.26. For each observation, we created
good time intervals from level 1 event files using standard filtering
criteria, for example requiring the source to be at least 30\deg\ from
the Earth's limb, and removing intervals around entry into and exit
from the South Atlantic Anomaly (SAA). In all of our \nicer\
  analyses, we only include photons in the range of 0.8--8~keV. Due to
  the non-negligible hydrogen column density towards \src\ and its soft
  X-ray spectrum, the background emission dominates below
  $\sim$0.8~keV and above $\sim$8~keV, respectively. Finally, we
removed MPU1 data from observation 2598041001 due to a time stamp
anomaly, which occurred on July 8th during \nicer\ passage through
SAA. This anomaly did not affect any of our subsequent observations.

The \swift/XRT is a focusing CCD, sensitive to photons in the energy
range of $0.2-10$~keV \citep{burrows05SSRv:xrt}. All XRT observations
we consider in this Letter were taken in windowed timing (WT) mode,
which results in a 1D image with a time resolution of 1.7~ms
\citep{evans07AA}. We reduced the data using \texttt{XRTDAS} version
3.5.0. We extracted source events from each good time interval (GTI)
of a given observation separately, using a circular region with a
20-pixel radius centered on the brightest pixel of each 1D
image. We extract background events from an annulus centered at the
same position as the source with inner and outer radii of 80 and 120
pixel. For the spectral analysis we generated the ancillary files
using \texttt{xrtmkarf}, and used the response matrices in
\texttt{CALDB} v014. We excluded any GTI for which the source landed
within a 3 pixel distance from a bad column or the edge of the
CCD. The remaining spectra for each observation were added together,
along with the ancillary, background, and response files using the
HEASOFT tool \texttt{addspec}.

We only perform spectral analysis on the \swift\ data. We use XSPEC
version 12.10.1f \citep{arnaud96conf}. To account for absorption
towards the source, we use the T\"ubingen-Boulder interstellar medium
absorption model  (\texttt{tbabs}) along with the abundances of
\citet{wilms00ApJ} and the photo-electric cross-sections of
\citet{verner96ApJ:crossSect}. We group the spectra to have at least
one count per spectral bin and use the Cash statistic (C-stat) in
XSPEC for model parameter estimation and error
calculation. We note that the background around \src\ is
  dominated by the emission from the SNR G109.1$-$1.0 (CTB~109,
  considered the progenitor to \src), which increases in intensity
  with increasing distance away from the magnetar \citep[and peaks
  around 3\arcmin\ from the source location, e.g.,][]{sasaki04ApJ:2259}.
  Accordingly, our XRT background estimate, which incorporates part of the SNR up
  to $2$\arcmin\ away from the source, should be considered a
  conservative correction to the SNR contribution to the source
  flux. Nevertheless, this background is only few percent of the source
  flux within our XRT source extraction region, even when considering
  the 0.8--2~keV energy range \citep[e.g.,][]{patel01ApJ2259,sasaki04ApJ:2259}.

We refrain from performing spectral analysis with \nicer\ given that
it is not an imaging instrument: the background within the
30~arcmin$^2$ field of view requires detailed, non-trivial
modeling. Along with the sky background, there is an unknown
contribution from the supernova remnant that depends on the placement
of the source within the field of view. Instead, for \nicer\
observations, we rely on pulsed flux analysis to check for any
variability in the source brightness level.

In total, we analyzed 117 \swift/XRT observations and 30 \nicer\
observations covering the time range between 2013 January 20 and 2019
September 10. We quote the uncertainties of all spectral and timing
model parameters at the 68\% confidence level, unless otherwise
noted.

\section{Results}
\label{res}

\begin{figure*}[!th]
\begin{center}
\includegraphics[angle=0,width=0.48\textwidth]{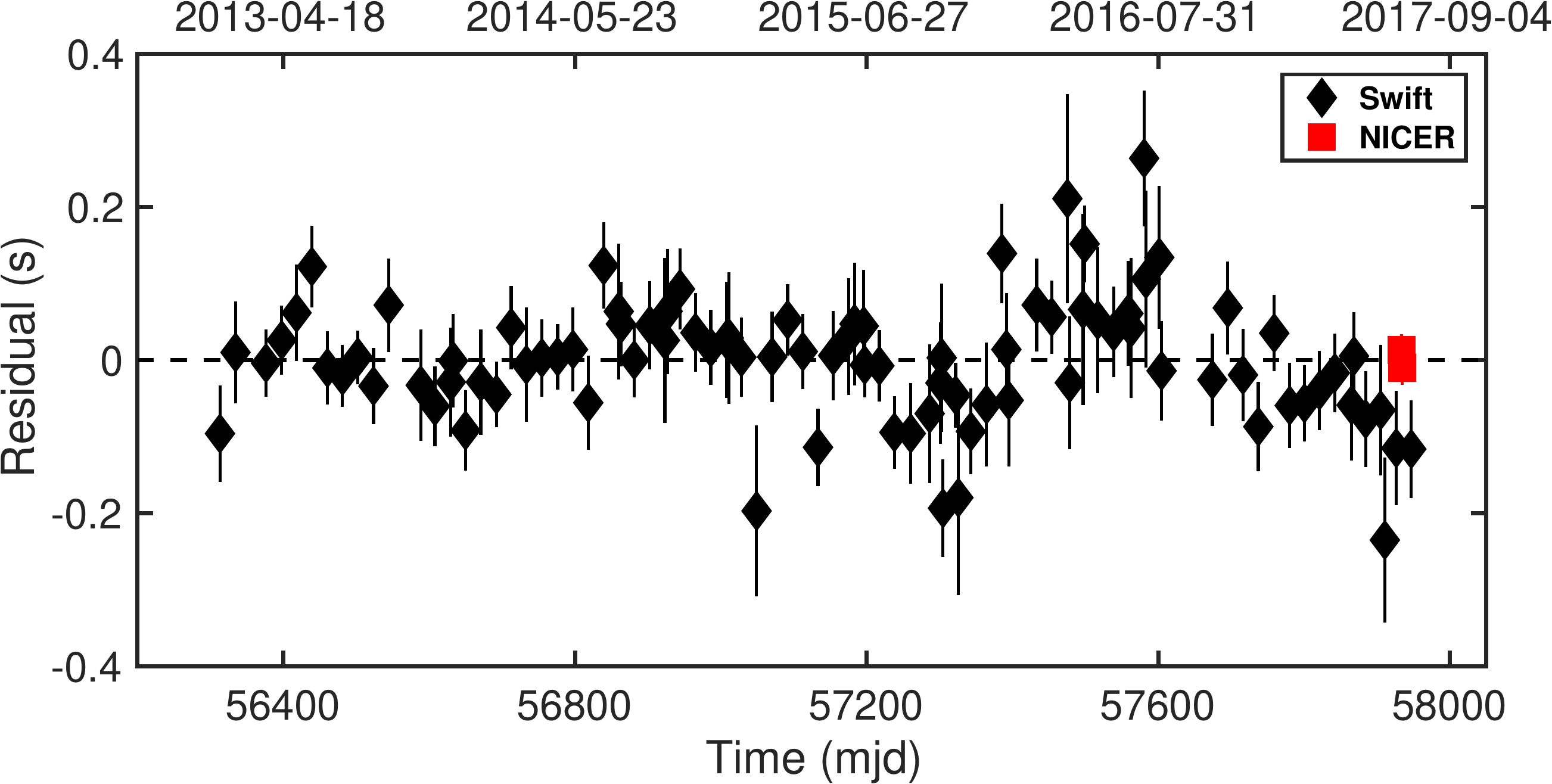}
\includegraphics[angle=0,width=0.486\textwidth]{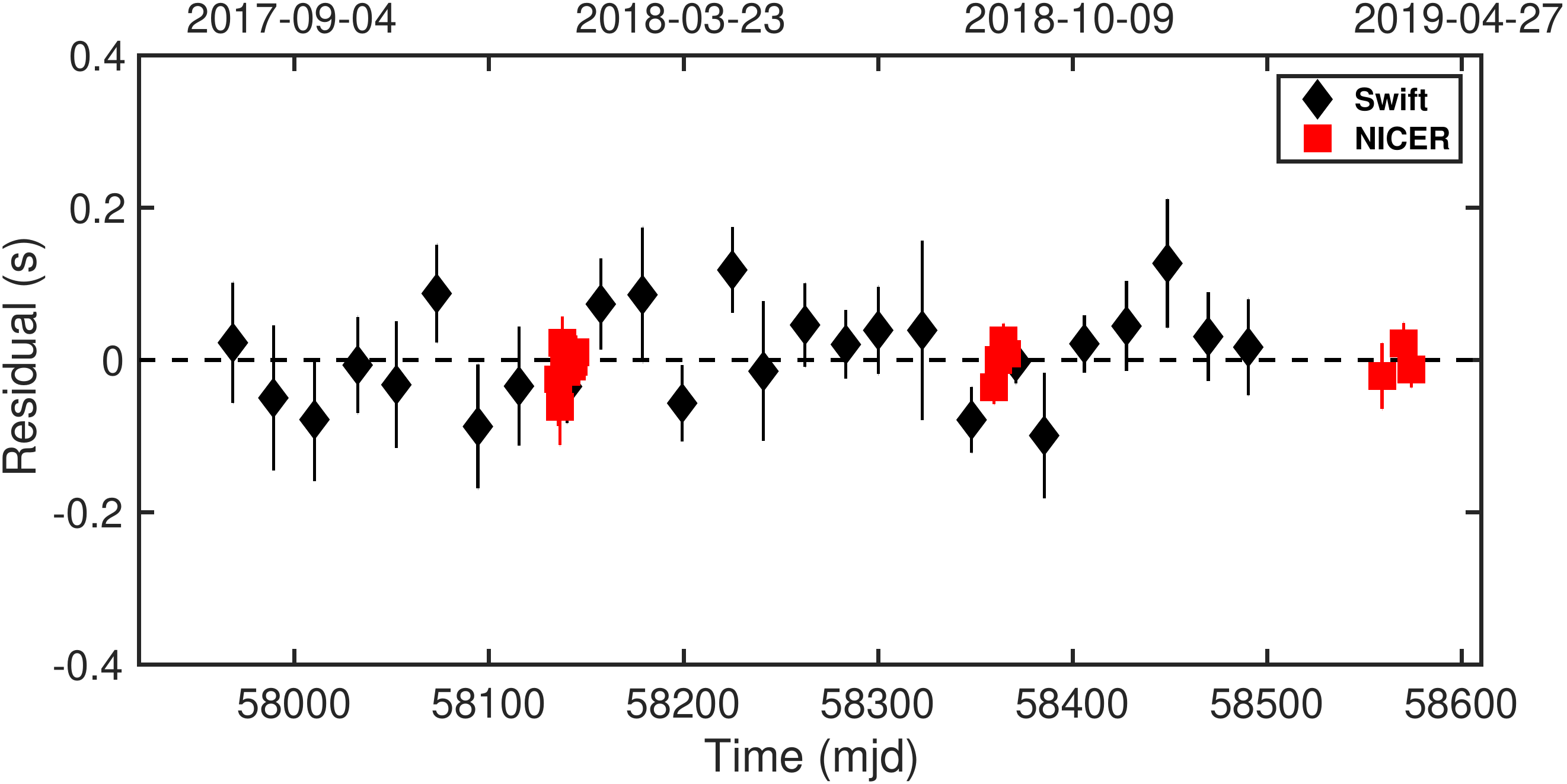}
\caption{Timing residuals for the two epochs in Table 1. {\sl Left
    panel.} Dates are from 2013 January to 2017 July. The best fit
  model includes contributions from the first 5 terms of
  equation~\ref{phShEq}. {\sl Right panel.} Dates are from 2017 August
  to 2019 April. Three frequency terms are included in the best fit
  model. See Table~\ref{timDat} for more details.}
\label{timResid}
\end{center}
\end{figure*}

\subsection{Timing}
\label{timAna}

We relied on a prior \swift\ magnetar monitoring program
\citep[e.g.,][]{archibald13Natur} to build a phase-coherent timing
solution for the source. We analyze here all WT mode \swift/XRT
observations of \src\ since 2013 January 20 (observation ID
00032035053); the first observation after the last one reported in
\citet{archibald13Natur}.  We selected photons in the energy range
$0.8-8$~keV (for consistency with \nicer; note that extending the
energy range to 10~keV does not have any impact on our timing
models), and corrected their arrival times to the solar barycenter
using the source best sky location \citep{Hulleman01ApJ}.
We then performed our phase-coherent timing analysis following a
phase-fitting technique \citep[e.g.,][]{dallosso03ApJ}. The source
pulse phase evolution is described by

\begin{equation}
  \label{phShEq}
  \phi(t)=\phi_0+\nu(t-t_0)+\frac{1}{2}\dot{\nu}(t-t_0)^2+\frac{1}{6}\ddot{\nu}(t-t_0)^3+\ldots,
\end{equation}

\noindent truncated to the highest statistically-significant term. We
first establish a spin period with high level of accuracy
utilizing several observations closely spaced in time. The phase drift
in such a select case is fit to the linear term of
equation~\ref{phShEq} and the spin frequency is corrected accordingly.
As more observations are added, the error on the spin frequency
decreases, until the phase drift is dominated by a spindown term. A
second term of equation~\ref{phShEq} is then added to the model and
the procedure is continued.

Following the method above, we were able to successfully phase-connect
all \swift\ observations from 2013 January 20 (MJD 56312) to 2017 July
  13 (MJD 57947), when our model failed to accurately predict the phase of the
subsequent observations, indicating the occurrence of a sudden timing
discontinuity. Our timing model spanning this date range required terms up to the fourth frequency derivative from equation~\ref{phShEq}. We
find a reduced chi-square, $\chi^2_\nu$, of 1.37 for 87 degrees of
freedom (dof) and an unweighted root-mean-square (rms) of 0.011
cycles. The best fit model parameters are summarized in
Table~\ref{timDat} while the residuals are shown in the left panel of
Figure~\ref{timResid}. Excluding the last term from our model results
in a slightly worse fit with a $\chi^2_\nu$ of 1.51 for 88 dof and an
rms of 0.013 cycles. We note that our timing model during
  this time-span predicts a spin frequency for the last observation
  reported in \citet{archibald13Natur} that is consistent at the
  $1\sigma$ and $3\sigma$ levels with their timing models 1 and 2,
  respectively.

%-----------
% Table  1
%-----------
\begin{table}
\caption{Phase coherent spin parameters of \src}
\label{timDat}
\vspace*{-0.5cm}
\begin{center}
\resizebox{0.48\textwidth}{!}{
\hspace*{-1.0cm}
\begin{tabular}{l c c}
\hline
\hline
MJD range \T\B &  56312--57947 & 57968--58574 \\
Epoch (MJD) \T\B & 57934.48 & 58359.56 \\
  \hline
  $\nu$ (Hz) \T\B &  0.143~282~728~7(3) & 0.143~282~545~6(2)\\
  $\dot\nu$ (Hz s$^{-1}$) \T\B &  $-$9.84(5)$\times10^{-15}$ & $-$9.75(2)$\times10^{-15}$\\
  $\ddot\nu$ (Hz s$^{-2}$) \T\B &  $-$1.8(3)$\times10^{-23}$ & 1.4(3)$\times10^{-23}$\\
  $d^3\nu/dt^3$ (Hz s$^{-3}$) \T\B & $-$6(1)$\times10^{-31}$ & $\ldots$\\
  $d^4\nu/dt^4$ (Hz s$^{-4}$) \T\B &  $-$6(1) $\times10^{-39}$ & $\ldots$\\
\hline
  $\chi^2$/dof \T\B &  117/87 & 31/39 \\
   RMS residual (cycle) \T\B &  0.011 & 0.0072 \\
  \hline
\hline
\end{tabular}}
\end{center}
%\vspace*{-0.2cm}
{\bf Notes.} The MJD ranges are for inter-glitch epochs.
\end{table}
%-----------
% Table  1
% -----------

To accurately describe the discontinuity that occurred after the 2017
July 13 observation, we focus our analysis on a time range centered on
this date and extending 9 months before and after the
anomaly\footnote{This baseline should suffice to properly constrain
  the anomaly parameters, search for any strong timing noise around
  this time, and constrain the presence of any
  exponentially-recovering frequency change.}. The upper-left panel of
Figure~\ref{glAntGl} shows the phase residuals for all observations after subtracting a
model consisting of $\nu$ and $\dot{\nu}$ as measured within the
9~month observation prior to the first anomaly epoch. The subsequent
drift in the pulse phase is linear in time indicating the presence of
a glitch dominated by a sudden change in the spin frequency. We fit
the full 18~month dataset with a model consisting of $\nu$,
$\dot{\nu}$, and a glitch model of the form

\begin{equation}
  \label{glEqu}
  \nu(t) = \nu_{\rm t}+\Delta\nu+\Delta{\dot{\nu}}(t-t_{\rm g}),
\end{equation}

\noindent where $\Delta\nu$ and $\Delta\dot{\nu}$ are the resultant
(semi-permanent) changes in spin frequency and its derivative, $t_{\rm
  g}$ is the glitch epoch, and $\nu_{\rm t}$ is the predicted spin
frequency prior to the glitch. We find a good fit to the phase-drifts
with a $\chi^2_\nu$ of 1.09 for 34 dof, and an unweighted rms of
$0.0092$ cycles. The middle-left panel of Figure~\ref{glAntGl} shows
the residuals in seconds of the best fit model. We find the
  glitch epoch $t_{\rm g}=57967.2(8)$~MJD, i.e., 2017 August 02. We
find a change in spin frequency $\Delta{\nu} = 1.78(2) \times
10^{-7}$~Hz~s$^{-1}$, and in spindown $\Delta\dot{\nu} = -3(2) \times
10^{-16}$~Hz~s$^{-2}$. We caution that this latter component is
required by the data only at the 2$\sigma$ level according to an
F-test and may not represent a true change in $\dot{\nu}$ but may
simply account for some timing noise present when considering long
stretches of the data. We note that the $\nu$ and $\dot\nu$
  we derive during this time-span is consistent within $2\sigma$ with
  the values derived in the first inter-glitch timing model
  (Table~\ref{timDat}, second column).

We built a phase-coherent timing model for the data starting
with the first observation after the glitch epoch, 2017 August 3 (57968~MJD), and
up to 2019 April 1 (58574~MJD). The phase drift in this time span is well fit with a
model consisting of three terms of equation~\ref{phShEq}. We find a
$\chi^2_\nu$ of 0.81 for 39 dof with an rms of 0.0072 cycles. The
timing solution that best describes the spin evolution of \src\ during
this time span is summarized in Table~\ref{timDat}, while the
residuals are shown in the right panel of Figure~\ref{timResid}.
However, this model does not successfully predict the pulse arrival
times of the \nicer\ observations after 2019 April 1, indicating the
detection of another anomaly.

Similar to our above method, we focus our phase-coherent timing
analysis around the time of the discontinuity. We include data
spanning 10 months prior to the 2019 April 1 observation, and up to
the 2019 September 10 \nicer\ observation (a total baseline
  of 16 months). A model consisting of a $\nu$ and $\dot{\nu}$
describes well the phase evolution of the source up to the time of
discontinuity. The phase-shifts according to this model are shown in
the upper-right panel of Figure~\ref{glAntGl}, where it is clear that
the true pulse arrival time of subsequent observations is lagging the
predicted one. Moreover, the phase shifts in all the following \nicer\
observations evolve linearly with time, implying a sudden, in this
case negative, jump in the spin frequency, i.e., the presence of an
anti-glitch.

We fit the full 16~months data set with a model consisting of $\nu$,
$\dot{\nu}$ and a sudden change in $\nu$, i.e., the first term of
equation \ref{glEqu}. We find a good fit to the data with a
$\chi^2_\nu$ of 1.27 for 26 dof and an rms of 0.0065 cycles. The
best fit sudden frequency change is $\Delta\nu=-8.3(1)\times
10^{-8}$~Hz~s$^{-1}$, with a fractional change $\Delta\nu/\nu=-5.8
\times10^{-7}$. Given the good observational coverage around the anti-glitch,
we constrain its epoch to within half a day, at 58574.5(5)~MJD, i.e., April 1, 2019.
The best fit parameters of the full model are summarized in
Table~\ref{glDat}, while the residuals in seconds are shown in the
middle-right panel of Figure~\ref{glAntGl}. Including a
sudden change in the frequency derivative at the time of the anti-glitch
does not improve the quality of the fit. Moreover, the
  $\nu$ and $\dot{\nu}$ derived through this model are consistent
  within $1\sigma$ with the ones derived in our second inter-glitch
  epoch (Table~\ref{timDat}, rightmost column). Finally, we note that
we do not detect any changes to the pulse profile shape in the
observations following either the spin-up or the spin-down glitch,
nor do we find any energy-dependent variability, hence,
there is no confusion in the timing model due to pulse counting.

\subsection{Spectroscopy}
\label{specAna}

\begin{figure*}
\begin{center}
\includegraphics[angle=0,width=0.50\textwidth]{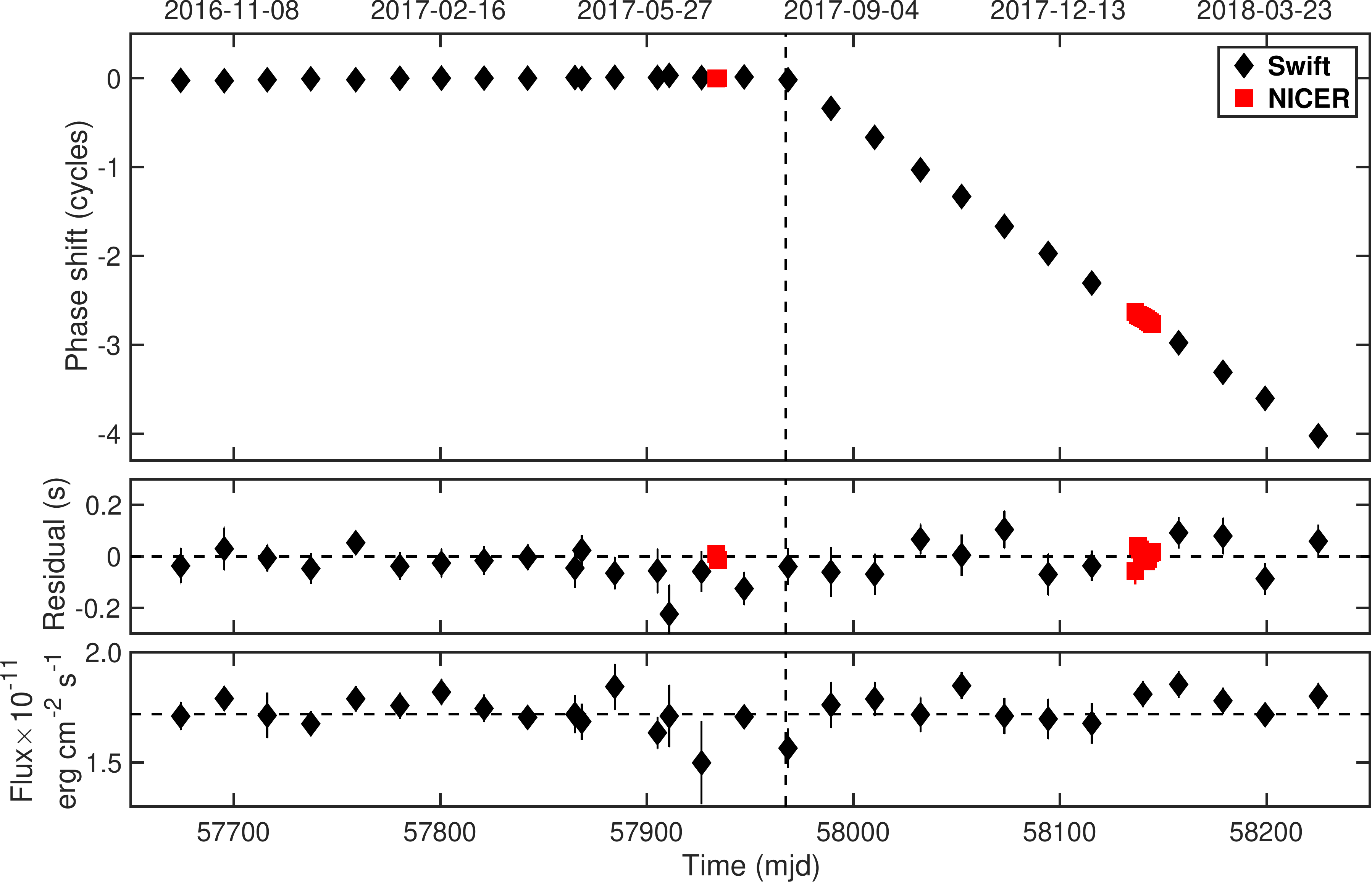}
\includegraphics[angle=0,width=0.49\textwidth]{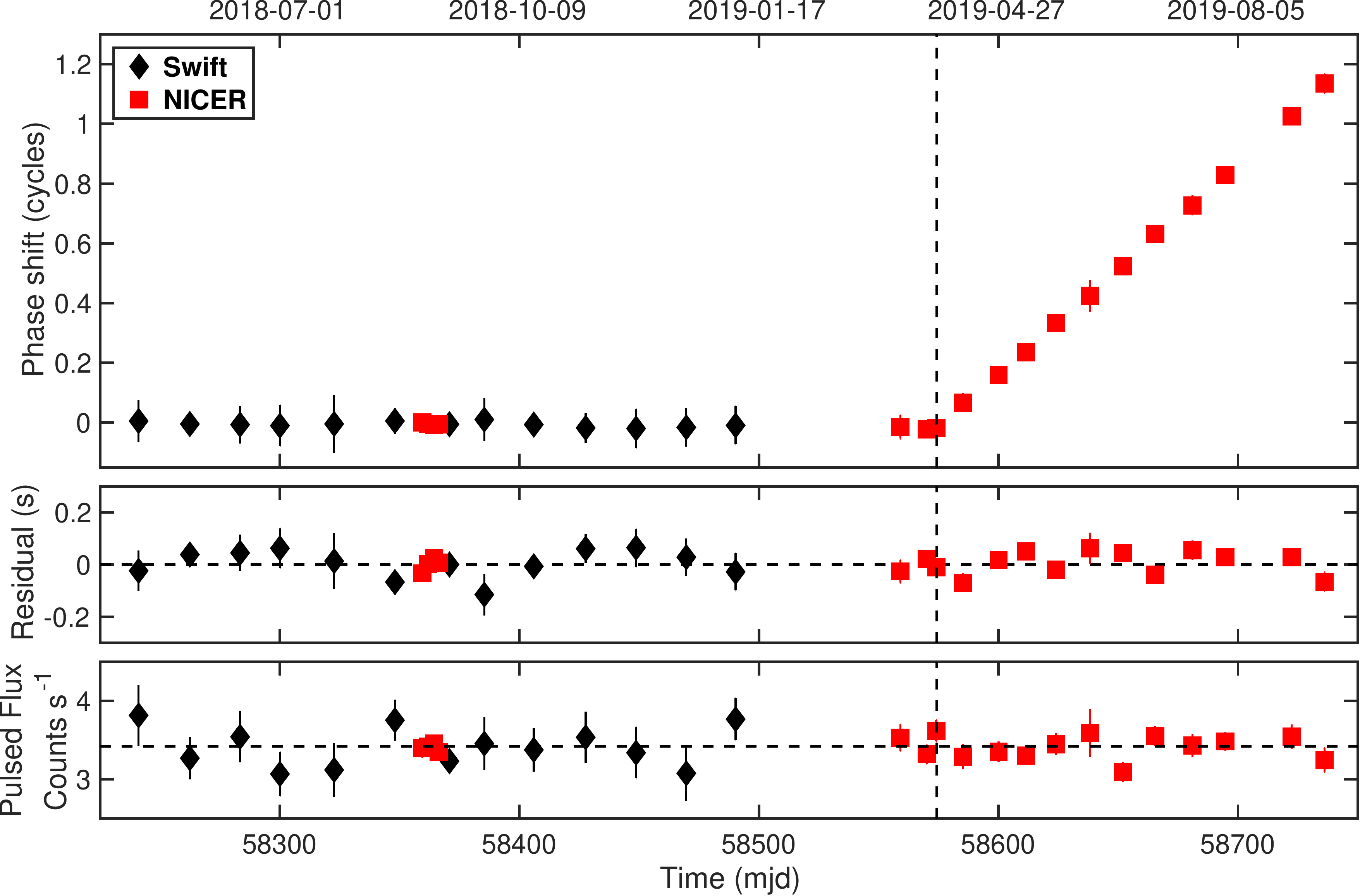}
\caption{{\sl Upper-left panel}. Timing residuals around the 2017 August
  glitch after subtracting a model consisting of $\nu$ and $\dot\nu$
  that best fit the pre-glitch data. {\sl Middle-left panel}.
  Timing residuals when including a glitch to the timing model
  (Table~\ref{glDat}, second row). The rms of the best fit is 0.0089
  cycles. {\sl Lower-left panel.} Absorption-corrected 2-10~keV flux as
  derived using \swift/XRT. The horizontal line is the average flux as
  derived with pre-glitch data. The vertical dotted line in all three
  panels represents the glitch epoch. {\sl Upper-right panel}. Timing
  residuals around the 2019 April glitch after subtracting a model
  consisting of $\nu$ and $\dot\nu$ that best fit the pre-glitch
  data. {\sl Middle-right panel}. Timing residuals when including a
  glitch to the timing model (Table~\ref{glDat}, third row). The rms
  of the best fit is 0.0065 cycles. {\sl Lower-right panel.} \nicer\
  0.8--8~keV rms pulsed flux. The horizontal dashed line is the average
  pulsed flux for the \nicer\ pre-glitch data. The \swift/XRT pulsed
  fluxes are normalized to this average. The vertical dotted line in
  all three panels represents the anti-glitch epoch. See text for more
  details.}
\label{glAntGl}
\end{center}
\end{figure*}

%-----------
% Table  2
%-----------
\begin{table}
\caption{Spin parameters around glitch epochs}
\label{glDat}
\vspace*{-0.5cm}
\begin{center}
\resizebox{0.48\textwidth}{!}{
\hspace*{-1.0cm}
\begin{tabular}{l c c}
\hline
\hline
MJD range \T\B &  57674--58225 & 58241--58736 \\
Epoch (MJD) \T\B & 57934.48 & 58359.56 \\
  \hline
  $\nu$ (Hz) \T\B & 0.143~282~729~9(9) & 0.143~282~546~2(6)\\
  $\dot\nu$ (Hz s$^{-1}$) \T\B &  -9.6(1)$\times10^{-15}$ & -9.74(6)$\times10^{-15}$\\
  $t_{\rm g}$ (MJD) \T\B & $57967.2(8)$ & $58574.5(5)$\\
  $\Delta\nu$ (Hz) \T\B &  $1.78(2)\times10^{-7}$ & $-8.3(1)\times10^{-8}$\\
  $\Delta\dot\nu$ (Hz~s$^{-1}$) \T\B & $-3(2)\times10^{-16}$ & \ldots\\
  $\Delta\nu/\nu$ \T\B & $1.24(2)\times10^{-6}$ & $-5.8(1)\times10^{-7}$\\
\hline
  $\chi^2$/dof \T\B &  33/34 & 33/26\\
   RMS residual (cycle) \T\B &  0.0089 & 0.0065\\
  \hline
\hline
\end{tabular}}
\end{center}
%\vspace*{-0.2cm}
{\bf Notes.} The MJD ranges encompass glitch epochs. The 1 $\sigma$
uncertainty on each parameter is given in parentheses.
\end{table}
%-----------
% Table  2
% -----------

To check for variability around the 2017 August spin-up glitch epoch, we
relied on the spectra of all prior \swift/XRT observations that were
employed to measure the glitch parameters (Figure~\ref{glAntGl}, left
panels). We simultaneously fit the 0.8--10~keV spectra with an absorbed 
blackbody (BB) plus power-law (PL) model. We link the hydrogen column
density between all observations and find $N_{\rm
  H}=(0.91\pm0.03)\times10^{22}$~cm$^{-2}$. The BB temperature $kT$ ranged between
0.40 and 0.44~keV with an average 1$\sigma$ uncertainty of about 0.02,
while we find a photon index $\Gamma$ between 2.6 and 3.4 and an
average 1$\sigma$ uncertainty of 0.5. These values are typical of
\src\ as inferred with, e.g., \xmm\ \citep{pizzocaro19AA}. The
2-10~keV absorption-corrected fluxes of these spectra are shown in the
lower-left panel of Figure~\ref{glAntGl}. The dashed horizontal line
is the average value and corresponds to $F_{\rm avg}=(1.7\pm0.1)
\times 10^{-11}$~erg~cm$^{-2}$~s$^{-1}$. We fit the subsequent \swift/XRT 
observations simultaneously with the same model, while fixing the
hydrogen column density to the value as derived from the pre-glitch
fit\footnote{Fixing $N_{\rm H}$  better constrains
  small flux variations that may be masked by allowing the
  column density to vary. We verified that allowing $N_{\rm H}$ to vary does
  not change any of our results.}. The absorption-corrected 2--10~keV
flux of these spectra are also shown in the lower-left panel of
Figure~\ref{glAntGl}. It is clear that these fluxes follow the
pre-glitch average flux well, implying that no flux variability
occurred at or shortly after the glitch epoch. The temperature of the
BB component and the PL indices were also within the uncertainties of
the values derived in the pre-glitch data. For completeness,
  we also verified that the RMS pulsed flux does not show any
  variability around the glitch epoch.

We relied on the RMS pulsed flux \citep[e.g.,][Section~\ref{obs}]{
  dib14ApJ} to search for any spectral changes around the anti-glitch.
The lower-right panel of Figure~\ref{glAntGl} shows the 0.8--8~keV RMS
pulsed flux (not background-corrected) of all \nicer\
observations that were used to characterize the anti-glitch timing
parameters. We also measured the 0.8--10~keV RMS pulsed flux of all
\swift/XRT observations, and multiplied the results by a constant
normalization of $F_{\rm rms,n}/F_{\rm rms,s}\approx16$, where $F_{\rm
  rms,n}$ and $F_{\rm rms,s}$ are the pre-glitch average of the
\nicer\ rms pulsed fluxes and the average of the \swift/XRT rms pulsed
fluxes, respectively. The \nicer\ post-glitch RMS pulsed fluxes follow
well the expected average as measured with the pre-glitch data (dashed
line, Figure~\ref{glAntGl}, lower-right panel) implying the absence of
pulsed flux variability at or following the time of the
anti-glitch. We note that the \swift\ spectra during this time-span
are consistent with the ones measured around the glitch epoch, and
with the long-term quiescent spectral properties of the source.

\subsection{Burst search}

We utilized \fermi-GBM to search for magnetar-like bursts $\pm5$ days
around the dates of the two discontinuities, using the
CTIME data type (0.256~s temporal resolution) in an energy range
10--100~keV. Our search algorithm is based on
\citet{gavriil04ApJ:1E2259}, which calculates the Poisson probability
for an event to be a random fluctuation around a background-corrected
mean, flagging any low probability events as possible
bursts. Throughout the 20 day period, we find 29 candidate bursts. In
order to determine if these events were related to \src, we used the
\fermi-GBM subthreshold search, referred to as the Targeted Search
\citep{goldstein19}, to follow-up these times and provide sky
localizations for those candidates. The Targeted Search utilizes
continuous time-tagged event (CTTE) data with 2 $\mu$s precision to
search $\pm$30 seconds around a given time.

We do not find any of the 29 burst candidates to be spatially
coincident at the 90\%\ level with the location of \src. The GBM
lower-limit for the detection of a magnetar-like burst in the energy
range 10--100~keV is $\sim1.0\times10^{-7}$~erg~s$^{-1}$~cm$^{-2}$
\citep[e.g.,][]{vanderhorst12ApJ:1550}, which would translate to a
luminosity of about $2.0\times10^{38}$~erg~s$^{-1}$ at the 3.2~kpc
distance of \src\ \citep{kothes12ApJ}. This is one order of magnitude
smaller than the luminosity of the GBM burst detected from \src\ at
the 2012 anti-glitch epoch \citep{foley12GCN}. Hence, we can exclude
the possibility of a similar short burst around the time of this
latest anti-glitch, as well as at the time of the 2017
  glitch, unless it occurred during GBM Earth-occulted periods, which
make up 20\% of the time. Also, we cannot exclude the possibility of
fainter short bursts akin to the ones detected from several magnetars
with \nustar\ \citep[e.g.,][]{an14ApJ:bursts,younes20ApJ:1708}.

\section{Summary and discussion}
\label{discuss}

In this paper, we have analyzed over 6 years of monitoring data of the
magnetar \src\ taken with \swift-XRT and \nicer, starting from the
\swift\ observation taken on 2013 January 20, and up to 2019
September 10. Following a relatively quiet period that lasted for about
five years after the 2012 anti-glitch, the source showed in August
2017 a large glitch, dominated by a sudden spin-up jump in
rotational frequency with a fractional change of the order of
$1.24(2)\times10^{-6}$ that exhibits no evidence of a ``healing'' recovery in its ephemeris. The glitch was not accompanied with any
spectral or temporal changes, and we did not detect any magnetar-like
bursts with \fermi-GBM from the direction of the source in a 10-day
interval around the glitch epoch down to a limiting flux of $\sim1.0\times10^{-7}$~erg~cm$^{-2}$~s$^{-1}$. 

Although a non-negligible fraction of magnetar glitches occur during
periods of outbursts \citep[e.g.,][]{archibald17ApJ:0142}, many are
observed in isolation, in the absence of any form of activity
\citep{dib14ApJ}. This is reminiscent of glitches observed from young
RPPs, where all glitch events (barring those detected from the two
high-B pulsars that showed magnetar-like activity, PSR~J1846$-$20,
\citealt{gavriil08Sci:psr1846}; and PSR~J1119$+$6127, \citealt{
  weltevrede11MNRAS:1119}) occurred ``silently'', without any
measurable change to their emission \citep{lyne00MNRAS,
  espinoza11MNRAS:glitches,yu13MNRAS}. Consequently, the origin of
impulsive spin-up glitches is thought to be internal, involving a
transfer of angular momentum and rotational kinetic energy from the
fast-rotating inner superfluid to the outer crust; these near-surface
layers slow down faster due to external magnetic dipole braking
torques. This picture also offers a plausible scenario for spin-up
glitches in magnetars.  Yet it does not {\it a priori} account for
radiative changes occurring in tandem with glitches.  Such activity,
when correlated with abrupt changes in the timing solution, likely
signal a physical connection between the zones of angular
momentum/energy transfer and magnetic field lines that thread the
crust through to the magnetosphere.  Mobility of heat/energy transfer
is enhanced along field lines in neutron stars, and thus a coupling of
surface and magnetospheric (burst) activity to spin-up glitches
suggests a concomitant threading of field lines deeper into the crust
proximate to the superfluid zones.

The timing solution following the spin-up glitch was robust enough to
predict the pulse arrival times up to April 1, 2019. The subsequent
observations indicate that a sudden spin-down glitch occurred, with
the pulse progressively lagging its predicted arrival time. The lag increased
monotonically with time implying a spin-down glitch dominated by a
change in rotational frequency. The fractional change of
this most recent anti-glitch is $-5.8(1)\times10^{-7}$, similar to the one detected in 2012 \citep{archibald13Natur,
  hu14ApJ:2259}. However, there are some notable differences 
between the two events. During the 2012 anti-glitch, \src\ entered an active period;
the 2--10~keV flux increased by a factor of 2, the spectrum exhibited
hardening, the shape of the pulse profile changed, and typical short
magnetar-like bursts  were detected from the direction of the source
with \fermi-GBM. On the other hand, we do not  detect any of the above
activity during the most recent anti-glitch. While short bursts could
have occurred during GBM earth-occulted periods or below its sensitivity level, our \nicer\
observations should have been able to detect any changes to the pulsed flux, pulse profile shape, and/or spectral properties of the source. This result shows that an
anti-glitch, similar to spin-up glitches, can indeed occur in
isolation, outside of outburst periods.

An interesting aspect of some of the magnetar and high-B pulsar
spin-up glitches detected at the onset of outburst activity (e.g., PSR
J1119$+$6127, \citealt{archibald18ApJ}, 4U~0142+61 \citealt{
  gavriil11ApJ,archibald17ApJ:0142}) is their over-recovery following
the glitch. Hence, the long-term effect on the rotational frequency is
a net spin-down. Assuming that the 2012 anti-glitch was due to an
over-recovery from a spin-up glitch, \citet{archibald13Natur} placed a
4 day $3\sigma$ upper-limit on the over-recovery time-scale for a
glitch size of $1.0\times10^{-6}$. Here, we place a slightly
  smaller $3\sigma$ upper-limit of 3~days for the recovery of a
glitch with the same size, which again is much smaller than
the typical weeks to months-long recovery usually observed
  for spin-up glitches.

Anti-glitches can be the result of interplay between
differentially-rotating portions of neutron stars, and/or adjustments to
the oblateness and moments of inertia of different regions. Extant
interpretations of anti-glitches center on two scenarios, although we
note that there is also the competing solid-body impact model of
\cite{huang14ApJ}. The first main paradigm is that adopted by
\cite{garcia15MNRAS} and \cite{mastrano15MNRAS} to address the 1E 2259+586 event in
2012, where adjustments in toroidal fields deep in the crust driven by
build up of magnetic tension lead to small but abrupt changes in the
overall oblateness of the star. The ``twisted torus'' field components
inflate the rotating star a little and yield metastable,
slightly-prolate configurations that over time reach a critical strain
that cracks the crust.  This irrepressible change re-organizes the internal
field to generate a slightly more spherical configuration of lower net
moment of inertia, yielding an impulsive frequency spin-up. The observed
$\vert \Delta\nu/\nu \vert \sim 3\times 10^{-7}$ suggested that the
toroidal fields are of the order of $\gtrsim 10^{15}$ Gauss in strength
(i.e., higher than the surface fields), and that the magnetic energy
released exceeded the observed radiative signal by a factor of $\sim5$.

The second picture is along the lines of more traditional models for
normal radio pulsar glitches in that the events are connected to vortex
reconfiguration/unpinning in the interior neutron superfluid zone but
close to the crust: see \cite{thompson00ApJ:1900} for a presentation
in the context of the giant flare from SGR 1900+14. \cite{kantar14ApJ}
discuss how the velocity difference between the superfluid and normal
stellar components can impact the degenerate energy and mass density
configuration.  They observe that vortex unpinning can actually
precipitate anti-glitches for some particular stellar differential
rotation profiles, yielding a deceleration in the rotation of both the
superfluid and crustal regions.  The associated adjustment of the radial
density profile and the moment of inertia of the superfluid is coupled
to an increase in the number of Cooper pairs. \cite{kantar14ApJ} find
that anti-glitches are more likely if the differential rotation between
core and crust is greater, and also if the superfluid core temperature
is larger, nominally $\gtrsim 5\times 10^7$~K.

In both scenarios, the site of crustal cracking and dissipation will
determine its connection to poloidal field lines and thereby develop a
geometric dichotomy for whether or not there would be an associated
energy release into the magnetosphere.  In the case of the 2019
``orphan'' anti-glitch event with no accompanying radiative enhancement, we suggest that the internal adjustment
locale is likely remote from the magnetic poles, and the reconfiguration energy
is deposited via heating of deep subsurface regions.

Yet, if the anti-glitch is accompanied by
plasma ejection into the magnetosphere, any increase in particle flux
along open field lines would raise torques on the star at the light
cylinder \citep{harding99ApJ:mag}, also contributing to a net 
spin-down. Since the rotationally-powered contribution to
the energetics of the magnetar is fractionally small, it is difficult to
calibrate any particle flux changes possibly associated with an
anti-glitch. Such enhanced winds could alter the flaring of the open field
line regions, possibly inducing changes in both the hard X-ray
persistent emission pulse profile and flux. Unfortunately, there
were no NuSTAR observations made in an epoch straddling the anti-glitch.
Thus a hard X-ray observational diagnostic on changes in the
magnetospheric wind properties is not afforded by this event.

Regardless of the origin and the physical mechanism triggering the
anti-glitches, \src\ must be unique in its internal structure. Our
detection of the spin-up and spin-down glitches from \src\ during the
last six years of monitoring demonstrates that the source undergoes
each timing discontinuity at a comparable rate, with the former having
a larger net fractional change. Throughout the full monitoring
campaign that was initiated with \rxte\ starting in 1997
\citep{kaspi03ApJ:2259}, \src\ has so far shown three spin-up glitches
and at least two spin-down glitches, and three if we include the
candidate 2009 event (we excluded the 2012 second timing discontinuity
here given that it could be interpreted as either a glitch or
anti-glitch). No other isolated neutron star has ever shown such a
sudden spin-down event, barring the disputed detection in the magnetar
1E~1841$-$045 \citep{sasmaz14MNRAS:1841,dib14ApJ}.
  
Compared to the rest of the magnetar population, \src\ is an archetypal
source. Its outburst activity is representative of the population; it
underwent two typical magnetar outbursts in the last $\sim25$ years
\citep{kaspi03ApJ:2259,archibald13Natur}. In quiescence, its broadband
X-ray spectrum is well described by a quasi-thermal soft X-ray part
and a hard X-ray tail, similar to almost all magnetars \citep[e.g.,][]{
  kuiper06ApJ,younes17ApJ:1806,enoto17ApJS}, while its timing
properties  are relatively stable, following the expected trend of
lower timing noise with increasing spin-down age \citep[e.g.,][]{
  cerri19MNRAS}. Hence, it is not clear what allows \src\ to undergo
anti-glitch events compared to other magnetars, and at such a high
rate. One noteworthy source in this regard is the accreting
ultra-luminous X-ray source NGC 300 ULX1 which showed a number of
anti-glitches during its most recent outburst
\citep{ray19ApJ}. However, in that case the anti-glitches came in the
context of a neutron star being spun up extremely rapidly, so an
anti-glitch is the natural consequence of the superfluid interior
lagging in that spinup. This cannot be the mechanism at work for
\src{} given that it is spinning down relatively consistently.
Continuing to monitor \src\ and the other bright magnetars is 
critical to better understand the causes of these distinctive events.

\section*{Acknowledgments}

We thank for the referee for comments helpful to the polishing of the
manuscript. GY acknowledges support from NASA under \nicer\ Guest Observer cycle-1
program 2098, grant number 80NSSC19K1452. M.G.B. acknowledges the
generous support of the NSF through grant AST-1813649. NICER work at
NRL is supported by NASA.

\end{document}